\documentclass[a4paper,12pt]{article}
\usepackage{amsmath}
\usepackage{epsfig}

\begin{document}
\title{Study Pure Annihilation Decays $B_s^0(\bar B_s^0)\to D^{\pm}\pi^{\mp}$ in
PQCD Approach\thanks{ This work is partly supported by National
Science Foundation of China.}
 }

\author{Ying Li\footnote{e-mail: liying@mail.ihep.ac.cn}, Cai-Dian L\"u\\
{\it \small CCAST (World Laboratory), P.O. Box 8730,
Beijing 100080, China}\\
 {\it \small  Institute of High Energy Physics,
CAS, P.O.Box 918(4) Beijing 100049, China}}
 \maketitle
\begin{abstract}
 The rare decays
$B_s^0\to D^\pm \pi^\mp$ and $\bar B_s^0\to D^\mp \pi^\pm$ can
occur only via annihilation type diagrams in the standard model.
In this paper, we calculate branching ratios of these decays in
perturbative QCD approach ignoring soft final state interaction.
From our calculation, we find that their branching ratios are at
$\mathcal{O}(10^{-6})$ with large CP asymmetry, which may be
measured in LHC-b experiment in future.
\end{abstract}

\section{Introduction}\label{s1}

The rich data from two B factories make the study of B physics a
very hot topic. A lot of study has been made, especially for the
CP violation problem. The CKM angle $\beta =\phi_1$ has already
been measured \cite{beta}. However the other two angles are
difficult to measure in B factories. The study of $B_s$ meson
decay is needed for this purpose. Some work on $B_s$ decays have
already been done \cite{bs,bsanni}.

In this work, we will explore four decay channels, namely
$B_s^0\to D^\pm \pi^\mp$ and $\bar B_s^0\to D^\mp \pi^\pm$. There
is only one kind of contribution for each of the decay mode, thus
there is no direct CP violation for them. However there is still
CP violation induced by mixing, although they are decays with
charged final states (non CP eigenstates). They are quite
complicated since altogether four are involved simultaneously
\cite{lu}.

 From these decays, we find that the four quarks
in final states are different from the ones in $B_s^0$ meson. We
call this mode pure annihilation type decay. In the factorization
approach, this decay is described as $B_s^0$ annihilating into
vacuum and final states mesons produced from vacuum afterwards.
They are rare decays. Up to now, only PQCD approach can calculate
this kind of modes effectively. Using PQCD approach, we have
calculated many of this kind of decays \cite{bsanni,bdsk,pad}, and
some decays have been measured in B factory. Some information
about PQCD in detail can be found in ref.\cite{PQCD}.

\begin{figure}[htbp]
\begin{center}
\includegraphics[scale=0.7]{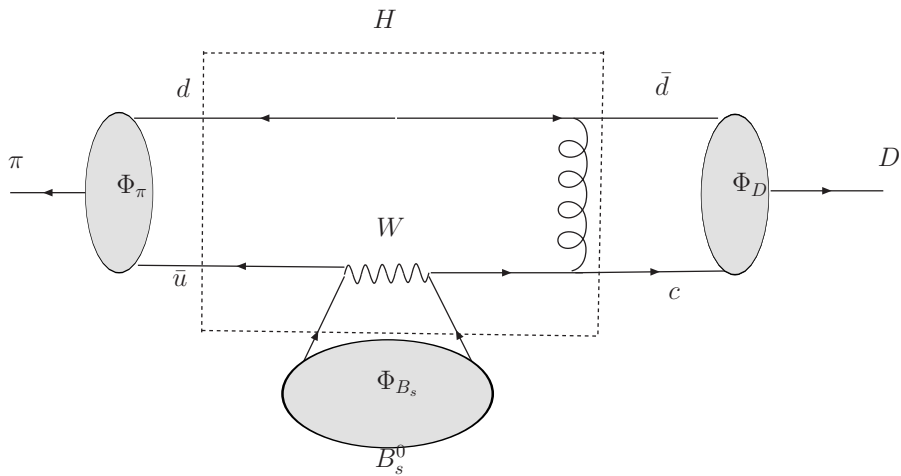}
\caption{The picture of PQCD approach.} \label{fig0}
\end{center}
\end{figure}

In standard model language, for decay $B_s\to D \pi$, a $W$ boson
exchange causes $\bar b s\to \bar u c$, and $\bar d d$ in final
state are produced from a gluon. This is also called $W$ exchange
diagram. This gluon can be emitted by any one of quarks
participating in the four quarks interaction. This is shown in
Figure.\ref{fig0}. We consider the $B_s^0$ meson at rest for
simplicity. In this frame, this gluon has $\mathcal{O}(M_B/2)$
momenta, that's to say, this is a hard gluon. We can
perturbatively treat the process by six quarks interaction.

In this work, we will give the PQCD calculation of these two
decays in the next section, and discuss the numerical results in
section \ref{s3}. At last we conclude this study in section
\ref{s4}.

\section{Calculation}\label{s2}

 The non-leptonic $B_s^0$ decays $B_s^0\to D^+ \pi^-$ and $B_s^0\to D^-
 \pi^+$ are rare decays. For decay $B_s^0\to D^+ \pi^-$, the effective Hamiltonian at the
scale lower than $M_W$ is given \cite{Buras} as:
\begin{eqnarray}
{\cal H}_1 = {G_F\over\sqrt{2}}\, V_{ub}^*V_{cs}
\Big[C_1(\mu)O_1(\mu)+C_2(\mu)O_2(\mu)\Big]\;,\label{h1}
\end{eqnarray}
where the four-quark operators are
\begin{eqnarray}
O_1= (\bar b s )_{V-A}(\bar c u)_{V-A}\;,\qquad\qquad O_2= (\bar b
u)_{V-A}(\bar c s )_{V-A}\;,
\end{eqnarray}
with the definition $(\bar q_1q_2)_{V-A}\equiv \bar
q_1\gamma_\mu(1- \gamma_5)q_2$.  $C_{1,2}$ are Wilson coefficients
at renormalization scale $\mu$. For decay $B_s^0\to D^- \pi^+$,
the effective Hamiltonian read:
\begin{eqnarray}
{\cal H}_2 = {G_F\over\sqrt{2}}\,V_{cb}^* V_{us}
\Big[C^{\prime}_1(\mu)O_1^{\prime}(\mu)+C_2^{\prime}(\mu)O_2^{\prime}(\mu)\Big]\;,\label{hf2}
\end{eqnarray}
where the four-quark operators are
\begin{eqnarray}
O_1^{\prime}= (\bar  b s)_{V-A}(\bar u c )_{V-A}\;,\qquad\qquad
O_2^{\prime}= (\bar
 b c)_{V-A}(\bar u s)_{V-A}\;.
\end{eqnarray}

In PQCD, the decay amplitude is expressed as \cite{PQCD}:
\begin{eqnarray}
 \mbox{Amplitude}
\sim \int\!\! d^4k_1 d^4k_2 d^4k_3\ \mathrm{Tr} \bigl[ C(t)
\Phi_{B_s^0}(k_1) \Phi_D(k_2) \Phi_\pi(k_3) H(k_1,k_2,k_3, t)
e^{-S(t)}\bigr]. \label{eq:convolution1}
\end{eqnarray}
In this Equation, $C(t)$ is Wilson coefficient at scale $t$ with
leading order QCD correction. $\Phi_i$ are light-cone wave
functions, which describe the non-perturbative contributions. They
can not be theoretically calculated directly. Fortunately, they
are process independent. $e^{-S(t)}$ is called Sudakov factor,
which comes from the resummation of soft and collinear divergence.
This Sudakov factor suppress the soft contributions, which make
the perturbative calculation of hard part reliable. By including
the $k_T$ dependence of the wave functions and Sudakov form
factor, this approach is free of endpoint singularity. Thus, the
work left is calculating the perturbative hard part $H(t)$.

The structures of the meson wave functions are
 \begin{eqnarray}
  B_s^0(P): &&[\not P + m_B ]\gamma_5 \phi_B(x)\;,\\
  D(P):  &&\gamma_5[\not P + m_D ]\phi_D(x)\;,\\
  \pi(P):&&
 \gamma_5 [\not P \phi_A(x) + m_0 \phi_P(x)
      + m_0(\not v\not n -1)\phi_T(x)]\;,
 \end{eqnarray}
 with $m_0 \equiv m_{\pi}^2/(m_u + m_d)=1.4$ GeV, characterizing the chiral breaking scale.
  And the light-like vectors are
 defined as $n=(1,0,{\bf 0}_T)$ and
 $v = (0,1,{\bf 0}_T)$. In above functions, $\phi_i$ are distribution
 amplitude wave functions.

\begin{figure}[htbp]
\vspace{-0.5cm}
\begin{center}
\includegraphics[scale=0.7]{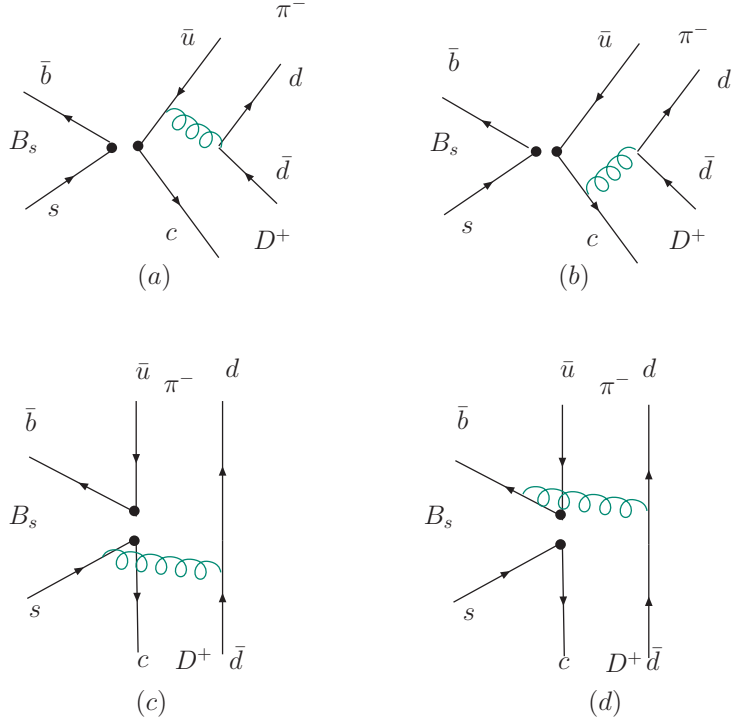}
\caption{Leading order Feynman diagrams contributing to decay
$B_s^0 \to D^+ \pi^-$.} \label{fig1}
\end{center}
\end{figure}

 According to the effective Hamiltonian (\ref{h1}), the diagrams
 contributing to $B_s^0\to D^+ \pi^-$ are drawn in Fig.\ref{fig1}. Just as
 stated in Section \ref{s1}, this decay has only annihilation
 diagrams. With the meson wave functions and Sudakov factors, the hard amplitude for
 factorizable annihilation diagrams in Fig.\ref{fig1}(a) and (b) is,
\begin{multline}
F_a= \frac{64\pi}{3} M_B^2 \int_0^1\!\!\! dx_2 dx_3
 \int_0^\infty\!\!\!\!\!  b_2 db_2\, b_3 db_3\ \phi_{D}(x_2,b_2)
\times \Bigl[ \bigl\{
\left( 1-x_3 \right) \phi_{\pi}^A(x_3) \\
+ r \left( 3 - 2 x_3 \right) r_{\pi} \phi_{\pi}^P(x_3)-r (1 - 2
x_3 ) r_{\pi} \phi_{\pi}^T(x_3)
\bigr\} E_{f}(t_a^1) h_a(x_2,x_3,b_2,b_3) \\
- \bigl\{ x_2\phi_{\pi}^A(x_3) + 2 r (1+x_2) r_{\pi}
\phi_{\pi}^P(x_3) \bigr\} E_{f}(t_a^2) h_a(1-x_3,1-x_2,b_3,b_2)
\Bigr], \label{F1a}
\end{multline}
where $r=m_D/M_{B_s^0}$, $r_{\pi} = m_{0}/M_{B_s^0}$, and the
functions $E_{f}$ containing Sudakov factors and Wilson
coefficients of four quark operator, hard scale $t_a^{1,2}$ and
virtual quark and gluon propagator $h_a$ are given in the
appendix. The explicit form for the distribution
 amplitude $\phi_M$ of wave functions,  are given in
the next section. In above function, $x_i$ denotes light (anti-)
quark momentum fraction in meson.

For the non-factorizable annihilation diagrams in
Fig.\ref{fig1}(c) and (d), the results can read:
\begin{multline}
M_a =  \frac{256\pi}{3\sqrt{2N_c}}   M_B^2 \int_0^1\!\!\! dx_1
dx_2 dx_3
 \int_0^\infty\!\!\!\!\! b_1 db_1\, b_2 db_2\
 \phi_{B_s^0}(x_1,b_1) \phi_{D}(x_2,b_2)\\
\times \Bigl[ \bigl\{ x_2 \phi_{\pi}^A(x_3,b_2)
 + r \left(1+x_2-x_3 \right) r_{\pi} \phi_{\pi}^P(x_3,b_2)\\
 + r \left(1-x_2-x_3\right) r_{\pi} \phi_{\pi}^T(x_3,b_2)
\bigr\}
E_{m}(t_{m}^1) h_a^{(1)}(x_1, x_2,x_3,b_1,b_2) \\
- \bigl\{
 \left( 1-x_3\right) \phi_{\pi}^A(x_3,b_2)
 + r \left(3 + x_2 -x_3\right) r_{\pi} \phi_{\pi}^P(x_3,b_2) \\
 + r \left(x_2 - 1 + x_3 \right) r_{\pi} \phi_{\pi}^T(x_3,b_2)
\bigr\}E_{m}(t_{m}^2) h_a^{(2)}(x_1, x_2,x_3,b_1,b_2) \Bigr].
\label{eq:Ma2}
\end{multline}

The total decay amplitude for $B_s^0\to D^+ \pi^-$ is given as
\begin{equation}
 A = f_B F_a + M_a . \label{eq:neut_amp}
\end{equation}
The decay width is expressed as
\begin{equation}
 \Gamma(B_s^0\to D^+ \pi^-) = \frac{G_F^2 M_B^3}{128\pi} (1-r^2)
|V_{ub}^*V_{cs} A|^2. \label{eq:neut_width}
\end{equation}

\begin{figure}[htbp]
\vspace{-0.5cm}
\begin{center}
\includegraphics[scale=0.7]{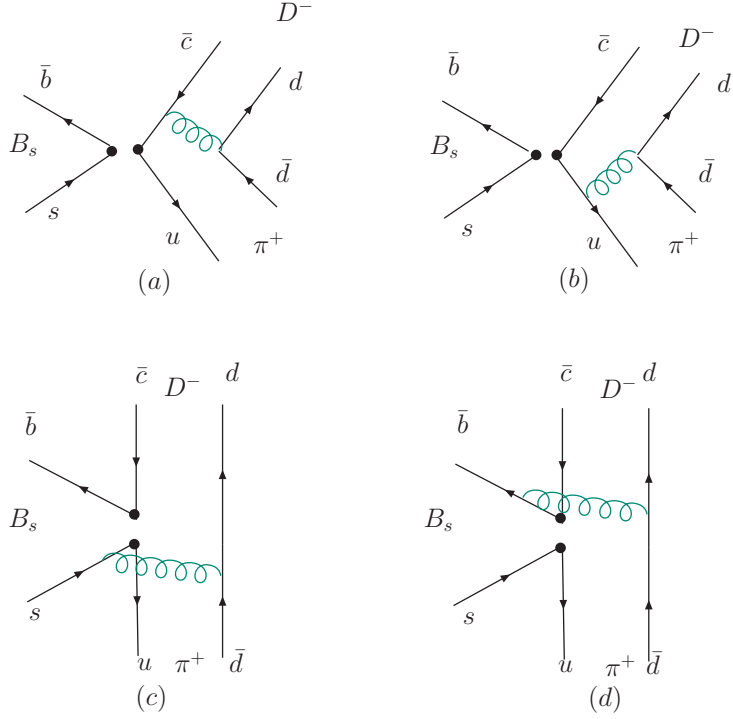}
\caption{Leading order Feynman diagrams contributing to decay
$B_s^0 \to D^- \pi^+$.} \label{fig2}
\end{center}
\end{figure}

As the case $B_s^0\to D^- \pi^+$, we also draw diagrams
Fig.\ref{fig2} using Equation (\ref{hf2}). The amplitude for
factorizable annihilation diagrams (a) and (b) results in $-F_a$.
 The amplitude for
the non-factorizable annihilation diagram results in
\begin{multline}
M_a^{\prime}  =  \frac{256\pi}{3\sqrt{2N_c}}   M_B^2
\int_0^1\!\!\! dx_1 dx_2 dx_3
 \int_0^\infty\!\!\!\!\! b_1 db_1\, b_2 db_2\
\phi_{B_s^0}(x_1,b_1) \phi_{D}(x_2,b_2)\\
 \times \Bigl[ \bigl\{
\left(1 - x_3  \right) \phi_{\pi}^A(x_3,b_2)
 + r \left(x_2 + 1-x_3 \right) r_{\pi} \phi_{\pi}^P(x_3,b_2)\\
 + r \left(x_2 - 1+x_3 \right) r_{\pi} \phi_{\pi}^T(x_3,b_2)
\bigr\}
E_{m}(t_{m}^1) h_a^{(1)}(x_1, x_2,x_3,b_1,b_2) \\
- \bigl\{  x_2 \phi_{\pi}^A(x_3,b_2)
 + r (3 + x_2 -x_3) r_{\pi} \phi_{\pi}^P(x_3,b_2)\\
 + r ( 1-x_2-x_3) r_{\pi} \phi_{\pi}^T(x_3,b_2)
\bigr\} E_{m}(t_{m}^2) h_a^{(2)}(x_1, x_2,x_3,b_1,b_2) \Bigr].
\label{Ma1}
\end{multline}

Thus, the total decay amplitude $A'$ and decay width $\Gamma$ for
$B_s^0\to D^- \pi^+$ decay is given as
\begin{gather}
  A' = -f_B F_a+ M_a^{\prime},
\label{eq:chrg_amp} \\
 \Gamma(B_s^0\to D^- \pi^+) = \frac{G_F^2 M_B^3}{128\pi} (1-r^2)
|V_{cb}^*V_{us} A'|^2 . \label{eq:chrg_width}
\end{gather}

The decays widths for CP conjugated mode, $\bar B_s^0\to
D^{\mp}\pi^{\pm}$, are the same expressions  as $B_s^0\to
D^{\pm}\pi^{\mp}$, with the conjugate of CKM matrix elements.

\section{Numerical Evaluation}\label{s3}

Considering SU(3) symmetry, we use the distribution amplitude of
the $B_s^0$ meson similar to $B$ meson:
\begin{equation}
\phi_{B_s^0}(x,b) = N x^2(1-x)^2 \exp \left[ -\frac{M_{B_s^0}^2\
x^2}{2 \omega_b^2} -\frac{1}{2} (\omega_b b)^2
\right],\label{waveb}
\end{equation}
which is adopted in ref.\cite{TLS,KLS,LUY}.  $N$ is a
normalization factor, which can be get from normalized relation:
\begin{equation}
 \int_0^1 \!\! dx\  \phi_M (x, b=0)
= \frac{f_M}{2\sqrt{2N_c}}. \label{eq:normalization}
\end{equation}
For $D$ meson, the distribution amplitude is
\begin{equation}
\phi_{D}(x) = \frac{3}{\sqrt{2 N_c}} f_{D} x(1-x)\{ 1 + a_{D} (1
-2x) \},
\end{equation}
which is fitted from experiments \cite{0305335}. The wave
functions of the $\pi$ meson have been derived in
ref.\cite{PB1,PB2}:
\begin{eqnarray}
\phi_\pi^A(x)&=&\frac{3f_\pi}{\sqrt{2N_c}} x(1-x)
\left[1+0.44C_2^{3/2}(2x-1)+0.25C_4^{3/2}(2x-1)\right]\;,
\label{pioa}\\
\phi_\pi^p(x)&=&\frac{f_\pi}{2\sqrt{2N_c}}
\left[1+0.43C_2^{1/2}(2x-1)+0.09C_4^{1/2}(2x-1)\right]\;,
\label{piob}\\
\phi_\pi^T(x)&=&\frac{f_\pi}{2\sqrt{2N_c}} (1-2x)
\left[1+0.55(10x^2-10x+1)\right]\;, \label{pioc}
\end{eqnarray}
with the Gegenbauer polynomials,
\begin{eqnarray}
& &C_2^{1/2}(t)=\frac{1}{2}(3t^2-1)\;,\;\;\;
C_4^{1/2}(t)=\frac{1}{8}(35 t^4 -30 t^2 +3)\;,
\nonumber\\
& &C_2^{3/2}(t)=\frac{3}{2}(5t^2-1)\;,\;\;\;
C_4^{3/2}(t)=\frac{15}{8}(21 t^4 -14 t^2 +1) \;.
\end{eqnarray}

The other input parameters are listed below \cite{pdg}:
\begin{eqnarray}
& & f_{B_s^0} = 230\; {\rm MeV}\;,\;\; \omega_B=0.5\;{\rm GeV}\;,
f_{D} = 240\; {\rm MeV}\;,\;\;C_D=0.8\pm 0.2\;,f_\pi = 132\;{\rm MeV}\;,\nonumber\\
& & m_B = 5.37\; {\rm GeV}\;,m_D = 1.87\; {\rm GeV}\;,m_0 = 1.4\;
{\rm GeV}\;, \tau_{B^0}=1.46\times
10^{-12}{\rm s}\;,\nonumber\\
&
&|V_{cb}|=0.043\;,\;\;|V_{us}|=0.22\;,\;\;|V_{ub}|=0.0036\;,\;\;|V_{cs}|=0.974\;.
\label{para}
\end{eqnarray}

With above parameters, we show the decay amplitudes calculated in
Table.{\ref{tb:amplitudes}}. The predicted branching ratios are:
\begin{gather}
 \mathrm{Br}(B_s^0 \to D^+ \pi^-) = 8.3\times 10^{-7},
\label{eq:br1}\\
 \mathrm{Br}(B_s^0 \to D^- \pi^+) = 2.9\times 10^{-6}.
\label{eq:br2}
\end{gather}

 From above results, we find that the branching ratios of decay
$B_s^0 \to D^+ \pi^-$ is  smaller than that of decay $B_s^0 \to D^-
\pi^+$. The CKM element in decay $B_s^0 \to D^+ \pi^-$ is
$V_{ub}^*V_{cs}$, but in $B_s^0 \to D^- \pi^+$ the CKM element is
$V_{cb}^*V_{us}$. Although $|V_{cb}^*V_{us}|$  and
$|V_{ub}^*V_{cs}|$ are both $\mathcal{O}(\lambda^3)$ in Wolfenstein
parametrization, the value of
$\frac{|V_{ub}^*V_{cs}|}{|V_{cb}^*V_{us}|}$ is   equal to 0.37. The
branching ratio of $B_s^0 \to D^- \pi^+$ is 3 times larger than that
of $B_s^0 \to D^+ \pi^-$, which is mainly due to the CKM factor.

 In
addition to the perturbative annihilation contributions, there are
another pictures existing such as soft final states interaction
\cite{fsi}. In ref.\cite{bdsk}, the results from the PQCD approach
for $B^0 \to D_s^-K^+$ is consistent with experiment well, which
tells us the soft final states interaction may not be important.
So we think their effects are small and ignore them in our
calculation.

\begin{table}[htbp]
 \begin{center}\caption{Decay amplitudes ($10^{-3}$ GeV) with parameters eqs.
 (\ref{F1a}-\ref{Ma1}).}
  \begin{tabular}[b]{cr|cr}
   \hline
   \hline
   \multicolumn{2}{c|}{$B_s^0 \to D^+\pi^-$} &
   \multicolumn{2}{c}{$ B_s^0 \to D^- \pi^+$} \\
   \hline \hline
   $f_B F_a$ & $0.51 -1.3\, i$ &
   $- f_BF_a$ & $-0.51 +1.3 \, i$ \\
   $M_a$ & $-16.1  -19.1\, i$ &
   $M_a^{\prime}$ & $-1.8 -19.1\, i$ \\
   $A$ & $-15.6 -20.4\, i$ &
   $A'$ & $-2.3 -17.8\, i$        \\
    \hline
   Br & $8.3 \times 10^{-7}$ &
   Br & $3.0 \times 10^{-6}$        \\
   \hline
   \hline
  \end{tabular}
 \end{center}
 \label{tb:amplitudes}
\end{table}

Unfortunately, there are no data for these two decays in
experimental side up to now. We think that the LHC-b experiment
can measure these decays in future. The results can test this PQCD
approach and show some information about new physics.

The calculated branching ratios in PQCD approach are sensitive to
various parameters such as the parameters in equations
(\ref{waveb}-\ref{para}). The uncertainty taken by $m_0$ has been
argued in many papers \cite{KLS,LUY}, and it is strictly
constrained by $B\to \pi$ form factor.  In
Table.\ref{tb:sensitivity}, we show the sensitivity of the
branching ratios to change of $B_s^0$ and $D$ distribution
amplitude  functions. It is found the uncertainty of the branching
ratio in PQCD is mainly due to $\omega_b$, which characterizes the
shape of $B_s^0$ meson wave function.
\begin{table}[htbp]
\caption{The sensitivity of the branching ratios to change of
$\omega_b$ and $a_D$} \label{tb:sensitivity}
\begin{center}
\begin{tabular}[t]{r|cc}
 \hline     \hline
& $(10^{-7})$ & $(10^{-6})$ \\
 $\omega_b$ & $\mathrm{Br}(B_s^0 \to D^+ \pi^-)$ &
$\mathrm{Br}(B_s^0 \to D^- \pi^+)$ \\
 \hline
 $0.45$ & $9.5$ & $3.5$ \\
 $0.50$ & $8.3$ & $2.9$ \\
 $0.55$ & $7.5$ & $2.5$ \\
 \hline
 \hline
 $a_D$ & $\mathrm{Br}(B_s^0 \to D^+ \pi^-)$ &
$\mathrm{Br}(B_s^0 \to D^- \pi^+)$ \\
 \hline
 $0.6$ & $7.6$ & $2.6$\\
 $0.8$ & $8.3$ & $2.9$\\
 $1.0$ & $9.1$ & $3.4$\\
 \hline
 \hline
\end{tabular}
\end{center}

\end{table}

Considering most of the uncertainty\footnote{Although the
uncertainty taken by CKM matrix elements is   large, we will not
discuss them in this work, since they are only an overall factor
here for branching ratios.}, we give the branching ratios of these
two decays with suitable range of $\omega_b$ and $a_D$. Thus we
can give our results:
\begin{gather}
\mathrm{Br}(B_s^0 \to D^+ \pi^-) = (8.3\pm^{1.2}_{0.8}) \times
10^{-7} \left( \frac{f_{B_s}\ f_{D}}{230\mbox{ MeV}\cdot 240\mbox{
MeV}} \right)^2 \left( \frac{|V_{ub}^*\ V_{cs}|} {0.0036\cdot
0.974} \right)^2
, \\
\mathrm{Br}(B_s^0 \to D^- \pi^+) = (2.9\pm^{0.5}_{0.5}) \times
10^{-6} \left( \frac{f_{B_s}\ f_{D}}{230\mbox{ MeV}\cdot 240\mbox{
MeV}} \right)^2 \left( \frac{|V_{cb}^*\ V_{us}|} { 0.0412 \cdot
0.224} \right)^2 .
\end{gather}

The $CP$ violation information in decay $B_s (\bar B_s)\to
D^{\pm}\pi^{\mp}$ is very complicated. There are four kinds of
decays
\begin{equation}\begin{array}{ll}
g=\langle D^+\pi^-|H|B_s^0\rangle ~\propto V_{ub}^*V_{cs} ,&
h=\langle D^+\pi^-|H|\bar B_s^0\rangle~\propto V_{cb}V_{us}^*,\\
\bar g=\langle D^-\pi^+|H|\bar B_s^0\rangle~\propto
V_{ub}V_{cs}^*, & \bar h=\langle D^-\pi^+|H|B_s^0\rangle~\propto
V_{cb}^*V_{us},\label{amp}
\end{array}\end{equation}
 which determine the decay matrix element of
$B_s^0\to D^+\pi^-$ and $D^-\pi^+$, and of $\bar B_s^0\to
D^-\pi^+$ and $D^+\pi^-$. They are already shown in the previous
Section. There is only one kind of contribution for each of the
decay mode, thus there is no direct CP violation for them. However
there is still CP violation induced by mixing, although they are
decays with charged final states \cite{lu}.

The   time-dependent decay rates for  $B_s  \to D^{\pm}\pi^{\mp}$
are given by:
\begin{eqnarray}
  \Gamma^{ D^\pm\pi^\mp} (t)
& =& (1\pm A_{CP}) \frac{e^{-t/\tau_{B_s}}}{8\tau_{B_s}} \left\{
1+  (S_{D\pi} \pm \Delta S_{D\pi}) \sin \Delta m t \right.\nonumber\\
&&\left. + (C_{D\pi} \pm \Delta C_{D\pi})\cos \Delta m t \right
\},
  \label{rate}
\end{eqnarray}
and $ \bar B_s \to D^{\pm}\pi^{\mp}$ by
\begin{eqnarray}
  \bar \Gamma^{ D^\pm\pi^\mp} (t)
& =& (1\pm A_{CP}) \frac{e^{-t/\tau_{B_s}}}{8\tau_{B_s}} \left\{
1- \left [ (S_{D\pi} \pm \Delta S_{D\pi}) \sin \Delta m t \right.\right.\nonumber\\
&&\left.\left. + (C_{D\pi} \pm \Delta C_{D\pi})\cos \Delta m t
\right] \right \}.
  \label{rate2}
\end{eqnarray}
Utilizing eq.(\ref{amp}), we can get
\begin{equation}\begin{array}{ll}
C_{D\pi} = A_{CP}=0 ,&
\Delta C_{D\pi} =\displaystyle\frac{ 1 -|h/g|^2}{ 1 +|h/g|^2},\\
 S_{D\pi} =\displaystyle \frac{2 \left|
 {h} /g\right| \sin \gamma \cos \delta} {1+|h/g |^2} ,&
 \Delta S_{D\pi}
=\displaystyle \frac{-2 \left|
 {h} /g\right| \sin \delta \cos \gamma} {1+|h/g |^2} .
\end{array}\label{aepsilon}\end{equation}
In deriving the above formulas we have neglected the small weak
phase $arg(q/p) =\frac{V_{tb}^* V_{ts}}{V_{tb}
V_{ts}^*}=2\lambda^2\eta< 2^\circ$, in Wolfenstein parametrization
\cite{wolfenstein}.

\begin{figure}[htbp]
\vspace{-0.5cm}
\begin{center}
\includegraphics[scale=1.3]{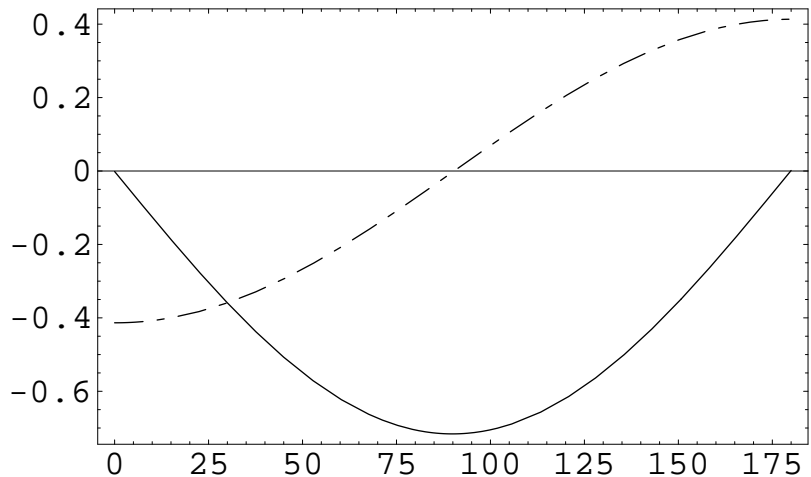}
\caption{CP violation parameters of $B_s^0(\bar B_s^0)\to
D^{\pm}\pi^{\mp}$: $\Delta S_{D\pi}$ (dash-dotted line) and
$S_{D\pi}$ (solid line) as a function of CKM angle $\gamma$.}
\label{fig4}
\end{center}
\end{figure}

We can calculate these parameters related to decays $B_s^0(\bar
B_s^0)\to D^{\pm}\pi^{\mp}$ in our PQCD approach. Through
calculation, we get:
\begin{equation}
\Delta C_{D\pi} = -0.56.
\end{equation}
The parameters $S_{D\pi}$ and $\Delta S_{D\pi}$ are $\gamma$
related. The results are shown in Figure.\ref{fig4}. If we can
measure the time-dependent spectrum of the decay rates of $B_s^0$
and $\bar B_s^0$, we can extract the CKM angle $\gamma $ and
strong phase $\delta $ in  eq.({\ref{aepsilon}}) by
Figure.\ref{fig4}. The parameter $C_{D\pi} =A_{CP}=0$ is from the
fact that there is only one kind of contribution to each of the
decays. Any deviation from zero, will be a signal of  new physics
contribution.

\section{Summary} \label{s4}

Recent study shows that PQCD approach works well for charmless B
decays \cite{KLS,LUY}, as well as for channels with one charmed
meson in the final states \cite{bdsk,0305335}. Because the final
state mesons are moving very fast,   each of them carrying more
than 2 GeV energy, there is not enough time for them to exchange
soft gluons. So we can ignore the soft final states interaction.
Due to   disadvantages in other approach such as general
factorization approach \cite{AGL} and BBNS approach \cite{BBNS},
pure annihilation decay can be calculated reliably only in PQCD
approach.

In this paper, we calculate $B_s^0 \to D \pi$ decays, which occur
purely via annihilation type diagrams. The branching ratios are
still sizable  at the order of $10 ^{-6}$. There will  also be
sizable CP violation in these decays. They will be measured in
future LHC-b experiment, which may bring  some information about
new physics to us.


\begin{appendix}

\section{Some functions}

The definitions of some functions used in the text are presented
in this appendix. In the numerical analysis we use one loop
expression for strong coupling constant,
\begin{equation}
 \alpha_s (\mu) = \frac{4 \pi}{\beta_0 \log (\mu^2 / \Lambda^2)},
\label{eq:alphas}
\end{equation}
where $\beta_0 = (33-2n_f)/3$ and $n_f$ is number of active flavor
at appropriate scale. $\Lambda$ is QCD scale, which we use as
$250$ MeV at $n_f=4$. We also use leading logarithms expressions
for Wilson coefficients $C_{1,2}$ presented in ref.\cite{Buras}.
Then, we put $m_t = 170$ GeV, $m_W = 80.2$ GeV, and $m_b = 4.8$
GeV.

The function $E_f^i$, $E_m$, and $E'_m$ including Wilson
coefficients are defined as
\begin{gather}
 E_{f}^i(t) = \big[C_1(t) +  {C_2(t)}{3}\big] \alpha_s(t)\, e^{-S_D(t)-S_{\pi}(t)}, \\
 E_{m}(t) = C_2(t) \alpha_s(t)\, e^{-S_B(t)-S_D(t)-S_{\pi}(t)}, \\
 E_{m}'(t) = C_1(t) \alpha_s(t)\, e^{-S_B(t)-S_D(t)-S_{\pi}(t)},
\end{gather}
 where
 $S_B$, $S_D$, and $S_{\pi}$ result from summing both double
logarithms caused by soft gluon corrections and single ones due to
the renormalization of ultra-violet divergence. The above $S_{B,
D, \pi}$ are defined as
\begin{gather}
S_B(t) = s(x_1P_1^+,b_1) +
2 \int_{1/b_1}^t \frac{d\mu'}{\mu'} \gamma_q(\mu'), \\
S_D(t) = s(x_2P_2^+,b_3) +
2 \int_{1/b_2}^t \frac{d\mu'}{\mu'} \gamma_q(\mu'), \\
S_{\pi}(t) = s(x_3P_3^+,b_3) + s((1-x_3)P_3^+,b_3) + 2
\int_{1/b_3}^t \frac{d\mu'}{\mu'} \gamma_q(\mu'),
\end{gather}
where $s(Q,b)$, so-called Sudakov factor, is given as
\cite{Li:1999kn}
\begin{eqnarray}
  s(Q,b) &=& \int_{1/b}^Q \!\! \frac{d\mu'}{\mu'} \left[
 \left\{ \frac{2}{3}(2 \gamma_E - 1 - \log 2) + C_F \log \frac{Q}{\mu'}
 \right\} \frac{\alpha_s(\mu')}{\pi} \right. \nonumber \\
& &  \left.+ \left\{ \frac{67}{9} - \frac{\pi^2}{3} -
\frac{10}{27} n_f
 + \frac{2}{3} \beta_0 \log \frac{\gamma_E}{2} \right\}
 \left( \frac{\alpha_s(\mu')}{\pi} \right)^2 \log \frac{Q}{\mu'}
 \right],
 \label{eq:SudakovExpress}
\end{eqnarray}
 $\gamma_E=0.57722\cdots$ is Euler constant,
and $\gamma_q = \alpha_s/\pi$ is the quark anomalous dimension.

The functions $h_a$, $h_a^{(1)}$, and $h_a^{(2)}$  in the decay
amplitudes consist of two parts: one is the jet function
$S_t(x_i)$ derived by the threshold resummation \cite{L3}, the
other is the propagator of virtual quark and gluon. They are
defined by
\begin{align}
& h_a(x_2,x_3,b_2,b_3) = S_t(1-x_3)\left( \frac{\pi i}{2}\right)^2
H_0^{(1)}(M_B\sqrt{(1-r^2)x_2(1-x_3)}\, b_2) \nonumber \\
&\times \left\{ H_0^{(1)}(M_B\sqrt{(1-r^2)(1-x_3)}\, b_2)
J_0(M_B\sqrt{(1-r^2)(1-x_3)}\, b_3) \theta(b_2 - b_3) + (b_2
\leftrightarrow b_3 ) \right\},
\label{eq:propagator1} \\
& h^{(j)}_a(x_1,x_2,x_3,b_1,b_2) = \nonumber \\
& \biggl\{ \frac{\pi i}{2}
\mathrm{H}_0^{(1)}(M_B\sqrt{(1-r^2)x_2(1-x_3)}\, b_1)
 \mathrm{J}_0(M_B\sqrt{(1-r^2)x_2(1-x_3)}\, b_2) \theta(b_1-b_2)
\nonumber \\
& \qquad\qquad\qquad\qquad + (b_1 \leftrightarrow b_2) \biggr\}
 \times\left(
\begin{matrix}
 \mathrm{K}_0(M_B F_{(j)} b_1), & \text{for}\quad F^2_{(j)}>0 \\
 \frac{\pi i}{2} \mathrm{H}_0^{(1)}(M_B\sqrt{|F^2_{(j)}|}\ b_1), &
 \text{for}\quad F^2_{(j)}<0
\end{matrix}\right),
\label{eq:propagator2}
\end{align}
where $\mathrm{H}_0^{(1)}(z) = \mathrm{J}_0(z) + i\,
\mathrm{Y}_0(z)$, and $F_{(j)}$s are defined by
\begin{equation}
 F^2_{(1)} = (1-r^2)(x_1 -x_2)(1- x_3),\
F^2_{(2)} = x_1 +x_2+(1-r^2)(1-x_1-x_2)(1-x_3).
\end{equation}
We adopt the parametrization for $S_t(x)$ of the factorizable
contributions,
\begin{equation}
 S_t(x) = \frac{2^{1+2c}\Gamma(3/2 +c)}{\sqrt{\pi} \Gamma(1+c)}
[x(1-x)]^c,\quad c = 0.3,
\end{equation}
which is proposed in ref.~\cite{Kurimoto:2001zj}. In the
non-factorizable annihilation contributions, $S_t(x)$ gives a very
small numerical effect to the amplitude. Therefore, we drop
$S_t(x)$ in $h_a^{(1)}$ and $h_a^{(2)}$. The hard scale $t$'s in
the amplitudes are taken as the largest energy scale in the $H$ to
kill the large logarithmic radiative corrections:
\begin{gather}
 t_a^1 = \mathrm{max}(M_B \sqrt{(1-r^2)(1-x_3)},1/b_2,1/b_3), \\
 t_a^2 = \mathrm{max}(M_B \sqrt{(1-r^2)x_2},1/b_2,1/b_3), \\
 t_{m}^j = \mathrm{max}(M_B \sqrt{|F^2_{(j)}|},
M_B \sqrt{(1-r^2)x_2(1-x_3) }, 1/b_1,1/b_2).
\end{gather}

\end{appendix}

\end{document}